\documentclass[prd,twocolumn,aps,showpacs,nofootinbib,nobibnotes,superscriptaddress,preprintnumbers]{revtex4-1}
\usepackage{epsfig}
\usepackage{graphicx}
\usepackage{bm}
\usepackage[usenames, dvipsnames]{color}
\usepackage{dcolumn}   
\usepackage[spanish,english]{babel}
\usepackage{bm}     
\usepackage{bbm}       
\usepackage{amssymb}  
\usepackage{amsmath}
\usepackage{latexsym}
\usepackage{float}
\usepackage{ifthen}
\usepackage{caption,subfig}
\usepackage{enumerate}
\usepackage{url}
\usepackage{caption,subfig}
\usepackage{amsopn}
\usepackage{hyperref}

\usepackage{amsfonts}
\usepackage{multirow}
\usepackage{array}
\usepackage{booktabs}
\usepackage{rotating}

\newcommand{\D}{\mathrm{d}}

\usepackage{ulem}
\def\clap#1{\hbox to 0pt{\hss#1\hss}}

\def\bea{\begin{eqnarray}}
\def\eea{\end{eqnarray}}
\def\be{\begin{equation}}
\def\ee{\end{equation}}
\def\mpl{M_{\rm P}}

\begin{document}

\title{Model Independent Analysis of Supernova Data, Dark Energy, Trans-Planckian Censorship and the Swampland}

\author{Lavinia Heisenberg} \email{lavinia.heisenberg@phys.ethz.ch}
\affiliation{Institute for Theoretical Physics,
ETH Zurich, Wolfgang-Pauli-Strasse 27, 8093, Zurich, Switzerland}
 
\author{Matthias Bartelmann} \email{bartelmann@uni-heidelberg.de}
\affiliation{Universit\"at Heidelberg, Zentrum f\"ur Astronomie, Institut f\"ur Theoretische Astrophysik, Germany}

\author{Robert Brandenberger} \email{rhb@hep.physics.mcgill.ca}
\affiliation{Physics Department, McGill University, Montreal, QC, H3A 2
T8, Canada}

\author{Alexandre Refregier} \email{alexandre.refregier@phys.ethz.ch}
\affiliation{Institute for Particle Physics and Astrophysics, Department of Physics, ETH Zurich, Wolfgang-Pauli-Strasse 27, 8093, Zurich, Switzerland}

\date{\today}

\begin{abstract}
In this Letter, we consider the model-independent reconstruction of the expansion and growth functions from the Pantheon supernova data. The method relies on developing the expansion function in terms of shifted Chebyshev polynomials and determining the coefficients of the polynomials by a maximum-likelihood fit to the data. Having obtained the expansion function in a model-independent way, we can then also determine the growth function without assuming a particular model. We then compare the results with the predictions of two classes of Dark Energy models, firstly a class of quintessence scalar field models consistent with the trans-Planckian censorship and swampland conjectures, and secondly a class of generalized Proca vector field models. We determine constraints on the parameters which appear in these models. 
\end{abstract}


\maketitle


\section{Introduction}

The origin of Dark Energy is one of the outstanding mysteries of contemporary physics. Could it be that the nature of Dark Energy could teach us something about quantum gravity?
Quantum gravity is a theory still in its infant form and under continuous development with the aim to unite quantum physics and general relativity. General relativity describes, at the level of classical physics, one of the four elementary forces of physics, namely gravitation. Quantum theory, on the other hand, is required to correctly describe the other three elementary forces, namely the electromagnetic, weak, and strong interactions. In order to consistently describe the interaction of quantum matter with gravity, a unification of all four forces at the quantum level is required. Einstein's theory of general relativity is still the best theory for the gravitational interaction below the Planck scale. So far, it describes in an exemplary manner all the empirical phenomena on a wide range of scales. Nevertheless, as a non-renormalizable theory, its UV completion into a quantum gravity theory remains a tenacious challenge. Even though quantum corrections in terms of higher order derivative self-couplings of curvature are Planck scale-suppressed, derivative couplings and self-interactions can induce infrared (IR) effects in many prominent gravity theories, like galileons, generalized Proca and massive gravity (see e.g. \cite{Heisenberg:2018vsk} for a recent review). In this sense, generalizations of general relativity are not just confined to the Planck scale but they do affect the IR behavior of the theory, and hence may be relevant for the Dark Energy mystery. Similarly, the emergence of curvature singularities in the form of Big Bang and Black Hole singularities in the realm of general relativity are disturbing properties of the theory. 

It is believed that general relativity could be the low energy limit of the more fundamental superstring theory. The latter represents a possible UV completion of gravity based on the idea that the fundamental building blocks of nature are strings with one-dimensional spatial extent instead of point-like elementary particles which underlie the usual quantum field theories. Since the 1980s, string theory is considered as the most promising candidate for a theory unifying all forces of nature at the quantum level, combining the standard model of elementary particle physics and gravitation. 
Should string theory be the ultimate theory of quantum gravity, an immediate question arises: how do the low energy effective field theories of gravity and matter known to us fit into string theory, and what are the constraints which string theory imposes on such effective field theories? In this context, the division of effective field theories into two groups, the ``Landscape'' and the ``Swampland'', has emerged (see e.g. \cite{Brennan, Palti} for reviews). Theories in the Landscape can successfully be embedded into a UV complete theory, whereas theories in the Swampland represent an inhabitable space incompatible with quantum gravity. For this purpose, a number of conjectures have been developed in order to delineate the division between Swampland and Landscape. Some evidence for these conjectures exists in stringy constructions but rigorous proofs are still lacking. Among them, the ``distance conjecture'' on the range of field values for fields which emerge from string theory moduli \cite{Vafa1}, the ``de-Sitter'' conjecture on the slope of the potential of such scalar fields \cite{Vafa2}, and the ``trans-Planckian censorship'' conjecture (TCC) \cite{BeVa} have received most attention (see also \cite{recent}). The criteria apply to the case when the scalar field (which is being constrained) dominates the energy density of the universe. This may have been the case in the early universe (e.g. in the context of inflation), but also in the late universe, in the phase of Dark Energy domination. Here, we will study the implications of the swampland conjectures for Dark Energy.

Both the de Sitter conjecture and the TCC imply that pure de Sitter space is incompatible with string theory, and lead to stringent constraints on early universe cosmology. For example, both imply dramatic constraints on a possible phase of early universe inflation (see e.g. \cite{VafaStein} for a study based on the de Sitter criterion, and \cite{BBLV} for an analysis based on the TCC) \footnote{Note that there are also quantum gravity arguments independent of string theory that prohibit de Sitter space \cite{Dvali}.}. The swampland conjectures explain why it has proven so difficult to construct stable de-Sitter vacua in string theory (see e.g. \cite{noLambda}, but see also \cite{yesLambda} for a different point of view).

Here, we are interested in exploring the connection between the swampland criteria and Dark Energy. As was mentioned above, the swampland conjectures are inconsistent with Dark Energy being a pure cosmological constant (the assumption which yields the ``vanilla'' $\Lambda$CDM model \footnote{The model which assumes that the Dark Energy is a cosmological constant, and that the dark matter is cold.}), and hence rule out what may be the simplest model of Dark Energy. Working in the context of Einstein gravity, the phase of late time acceleration could also be explained by invoking a slowly rolling scalar matter field, a ``quintessence field'' \cite{Peebles, Wetterich}. Constraints on quintessence models were already explored in \cite{VafaStein} and \cite{HBBR}. However, the analysis in those works was in the context of the standard cosmological model. Here, we revisit the constraints on Dark Energy using a model-independent determination \cite{BM2019} of the expansion and growth functions from the Pantheon supernova data sample \cite{Pantheon}.  
Specifically, we consider the class of quintessence models with an exponentially decaying potential \cite{Peebles, Wetterich}. The swampland criteria constrain the coefficient in the exponent. We find that given the current accuracy of the observations, there is already a difference between the predictions of the vanilla $\Lambda$CDM model and quintessence models which are preferred by the swampland arguments.

We also consider a class of Dark Energy models in which Dark Energy is a generalized Proca vector field. We find that the predictions for the expansion and growth functions differ in an interesting way. The Proca models which we study also contain a free parameter. We identify the range of this parameter for which the predictions of this model can be distinguished from that of the $\Lambda$CDM model at the current level of the observational accuracy. We find that the deviations in the expansion and growth functions in the Proca model (compared to those of the $\Lambda$CDM) have the opposite sign compared to those in quintessence models.

\section{Swampland Conjectures and Cosmology}

The Swampland criteria relevant for us concern effective scalar field theories canonically coupled to gravity and with a canonical kinetic term in a phase in which these fields dominate the energy density of the universe. For such theories to be consistent with string theory, conditions are (given a point in field space) \cite{Brennan, Palti}
\begin{itemize}  
\item The Swampland {\it distance conjecture}: the range traversed by a scalar field is bounded by $|\Delta\pi|<d\sim\mathcal{O}(1)$ in reduced Planck units \cite{Vafa1}, where $d$ is a positive constant of the order 1; 
\item the de-Sitter conjecture: the derivative of the scalar-field potential has to satisfy the lower bound $|V'|/V>c\sim\mathcal{O}(1)$ in reduced Planck units \cite{Vafa2} \footnote{The refined version of this conjecture states that if the condition on the first derivative is not satisfied, then a condition on the second derivative has to be fulfilled ${\rm min}(V'')/V\le-\tilde{c}\sim\mathcal{O}(1)$ \cite{Krishnan, Ooguri:2018wrx}. This is particularly relevant for models where the scalar field is close to a local maximum of the potential. Either of these conditions should be applied.} ; and
 \item the ``trans-Planckian censorship conjecture'': no perturbation in an expanding universe which ever had a length scale larger than the Hubble radius could have emerged from trans-Planckian scales at an earlier time. Mathematically, this can be expressed as $\frac{a_f}{a_i}<\frac{1}{H_f}$  \cite{BeVa}, with initial time $a_i$ and final time $a_f$, with $H_f^{-1}$ being the Hubble radius rate at the final time. This can be reformulated as $\int_{t_i}^{t_f}Hdt < \ln\frac{1}{H_f}$. Applied to exponential quintessence potentials, and extrapolating the evolution to infinite field values, this condition fixes $c$ in the de Sitter conjecture to be $|V'|/V>\frac{2}{\sqrt{(d-1)(d-2)}}$, which for 4 dimensions would mean $c=\sqrt{\frac23}$. However, in light of the distance conjecture it is unclear whether one can extrapolate the quintessence evolution in this way.
\end{itemize}
The distance conjecture states that if the scalar field traverses a larger distance than the one given by the bound, then one immediately leaves the domain of validity of the effective field theory as new string states become massless and a tower of light states pushes down the cutoff of the effective field theory. Within the cutoff domain of the effective field theory over which the field evolves, the second condition additionally constrains the slope of a positive potential. The trans-Planckian censorship conjecture then demands that no length scale which exits the Hubble horizon could ever have had a wavelength smaller than the Planck length, thus hiding the non-unitarity of the effective field theory from the domain where linear fluctuations can grow in amplitude (see \cite{Edward} for a discussion of this point). This leads to a duration limit of any phase of accelerated expansion. Hence, models with a de Sitter attractor would be in direct conflict with the TCC condition.

\section{Current observables, Future Evolution and the TCC}

In this work, we consider, within standard general relativity, an effective Dark Energy component which we will assume to be responsible for the late-time cosmic acceleration. The action of the theory is
\begin{equation}
\mathcal{S}=\int d^4x \sqrt{-g}\left\{ \frac{\mpl^2}{2}R+\mathcal{L}_{\rm DE}(\rho_{\rm DE},P_{\rm DE}) \right\}+\mathcal{S}_{\rm matter}\,,
\end{equation}
where $\mathcal{L}_{\rm DE}$ represents the Lagrangian of Dark Energy with energy density $\rho_{\rm DE}$ and pressure $P_{\rm DE}$ and $\mathcal{S}_{\rm matter}$ stands for the standard matter field. We also adopt the usual cosmological symmetries, i.e.\ spatial homogeneity and isotropy. Hence, the background metric will be of Friedmann-Lemaitre-Robertson-Walker (FLRW) form 
\begin{equation}
g_{\mu\nu}={\rm diag}(-N(t)^2,a(t)^2,a(t)^2,a(t)^2) \, ,
\end{equation}
and the Dark Energy and matter components will also be purely time dependent. The background evolves according to
\begin{eqnarray}\label{background_eom}
3\mpl^2H^2&=&\rho_{\rm DE}+\rho_m\,,\\
\mpl^2(3H^2+2\dot{H})&=&-P_{\rm DE}-P_m  \,.
\end{eqnarray}

The energy density and the pressure of the Dark Energy could originate from a scalar or a vector field, or any other field specific to the considered model, but at the background level they will most likely, under certain assumptions, manifest themselves in terms of an effective energy density and an effective pressure satisfying the background symmetries. At the background level, the combination of different cosmological observations allows us to determine the Hubble expansion rate, which can then be directly compared to the specific model predictions.

We want to compare such effective Dark Energy models with cosmological observables today and compare their future destiny with the condition imposed by the trans-Planckian censorship conjecture \cite{BeVa}. It translates into the condition
\begin{equation}
\frac{a_f}{a_i}<\frac{\mpl}{H_f}
\end{equation}
which is equivalent to
\begin{equation}\label{TCC}
\int_{t_i}^{t_f}H dt<\ln \left(\frac{\mpl}{H_f}\right)\,.
\end{equation}
Thus, on the one hand we have the Hubble expansion rate tightly constrained by a combination of different cosmological observations today, and on the other hand the upper bound \eqref{TCC} enforced by the trans-Planckian censorship conjecture only allows certain cosmic evolution in the future. In particular, de Sitter attractors are in direct disagreement with the TCC.

Another important observable is the growth rate, which requires the analysis of perturbations. The background metric including scalar fluctuations can be perturbed as (see e.g. \cite{MFB} for a review) 
\begin{equation}
d^2s \, = \, -(1-2\Phi)dt^2+a(t)^2(1+2\Phi)dx_i^2 \, ,
\end{equation}
where $\Phi$ is the relativistic gravitational potential. We are using longitudinal gauge and have assumed that there is no anisotropic stress at linear order.  Similarly, the Dark Energy and matter components can be perturbed. Introducing the matter density contrast $\delta_m=\delta\rho_m/\rho_m$, the dynamics of the linear matter perturbations satisfies
\begin{eqnarray}\label{growthEqQuint}
\delta_m''+\left(2+\frac{H'}{H}\right)\delta_m'-\left( \frac{k}{aH} \right)^2\Phi \nonumber\\
=-3\left(\Phi''+\left(2+\frac{H'}{H}\right) \right)\Phi' \,,
\end{eqnarray}
where we are working in Fourier space, $k$ is the comoving momentum and $'$ represents derivatives with respect to $N=\ln a$.

We are interested in the late time dynamics on scales which are large but sub-Hubble. In this case, the Poisson equation can be used to replace the $k^2 \Phi$ term by a function of $\delta_m$. In the quasi-static approximation the time dependence of $\Phi$ is negligible. Hence, on these scales the matter perturbations follow the approximate evolution equation
\begin{equation}\label{growthEqQuint}
\delta_m''+\frac12(1-2w\Omega_{\rm DE})\delta_m'-\frac32\Omega_m \delta_m\approx 0\,,
\end{equation}
where the matter density parameter satisfies $\Omega_m=1-\Omega_{\rm DE}$. The sub-Hubble approximation does not apply to modes with physical momenta $k/a$ sufficiently smaller than the Hubble expansion rate $H$, i.e. with $\epsilon=\frac{Ha}{k}\gg1$. The quasi-static approximation neglects the fast modes and keeps the slow modes in order to describe an adiabatic evolution.

\section{Model-Independent Reconstruction of the Hubble and Growth Functions}

In this section, we are interested in obtaining the constraints coming from cosmological observations. To be precise, we will construct the Hubble function from the Pantheon supernovae sample \cite{Pantheon} in a model-independent way, and subsequently obtain the growth function, following the analysis of \cite{BM2019}.

\subsection{Hubble function}

Two of the most important functions in cosmology are the Hubble function $H(a)$ and the linear growth function $D_+(a)$. For a homogeneous and isotropic background, characterized by the Friedmann-Lemaître-Robertson-Walker metric, the Hubble function is determined by the Hubble constant $H_0$ and the cosmic expansion function $E(a)$ by $H(a)=H_0E(a)$. In the standard $\Lambda$CDM model,
\begin{equation}
  E^2_{\Lambda\mathrm{CDM}}(a) =
  \Omega_{r0}a^{-4}+\Omega_{m0}a^{-3}+\Omega_{\Lambda0}+\Omega_Ka^{-2}
\end{equation} 
with the present-day density parameters of radiation, matter, cosmological constant and spatial curvature evaluated today, respectively.

In this work, we will however not adopt any specific form for $E(a)$ given by some cosmological model defined a priori. Rather, we use the expansion function reconstructed directly from suitable sets of measured distances to type-Ia supernovae (the Pantheon sample). We briefly review the reconstruction method and its results here; see \cite{BM2019} for details.

Given measurements $\{z_i, \mu_i\}$ of redshifts $z$ and distance moduli $\mu$ of individual supernovae (labelled by the index $i$), we transform the redshifts to scale factors $a_i = (1+z_i)^{-1}$ and these to
\begin{equation}
  x_i = \frac{a_i-a_\mathrm{min}}{1-a_\mathrm{min}}\;,
\end{equation}
with $a_\mathrm{min}:=\min(a_i)$, to map the scale factors to the standard interval $[0,1]$. We further convert the distance moduli to the luminosity distance by means of
\begin{equation}
  D_{\mathrm{lum},i} = 10^{1+0.2\mu_i}\,\mathrm{pc}
\end{equation} 
and introduce the scaled luminosity distance
\begin{equation}
  d_i := a_\mathrm{min}^2(1+\delta ax_i)D_{\mathrm{lum},i}
  \quad\mbox{with}\quad
  \delta a := \frac{1-a_\mathrm{min}}{a_\mathrm{min}}\;,
\end{equation} 
which is related to the cosmic expansion via
\begin{equation}
  d(x) = \int_x^1dx'\,e(x')\quad\mbox{with}\quad
  e(x) := \frac{\delta a}{E(a)(1+x\delta a)^2}\;.
\end{equation} 

In order to reconstruct the expansion function $E(a)$ in a model-independent way from the data set $\{x_i,d_i\}$, we now expand the function $e(x)$ into the shifted Chebyshev polynomials $T_n^*(x)$, which are orthonormal on the interval $[0,1]$ with the weight function $w^*(x):=(x-x^2)^{-1/2}$. The shifted Chebyshev polynomials are defined in terms of the Chebyshev polynomials via
\begin{equation}
  T_n(x) := \begin{cases}
              (1/\pi)^{1/2} & (n = 0) \\
              (2/\pi)^{1/2}\cos(n\arccos x) & (n > 0) \\
            \end{cases}
\end{equation}
by $T_n^*(x) = T_n(2x-1)$. We thus write the expansion rate in terms of these Chebyshev polynomials
\begin{equation}
  e(x) = \sum_{j=1}^Mc_jT_j^*(x)\;,
\end{equation} 
express the luminosity distance as
\begin{equation}
  d(x) = \sum_{j=1}^Mc_jP_j(x)\quad\mbox{with}\quad
  P_j(x) := \int_x^1dx'\,T_j^*(x)
\end{equation}
and find the coefficients $\vec c$ from the maximum-likelihood fit
\begin{equation}
  \vec c = \left(P^\top C^{-1}P\right)^{-1}\left(P^\top C^{-1}\right)\vec d
\end{equation} 
to the data. Here, $\vec d = (d_i)$ is the vector of measured, scaled luminosity distances, $C=\langle\vec d\otimes\vec d\,\rangle$ is its covariance matrix, and the matrix $P$ has the components $P_{ij} := P_j(x_i)$. From the Pantheon data, we obtain the $M = 2$ coefficients and their uncertainties listed in Tab.~\ref{tab:1}. Higher-order coefficients turn out to be insignificant in the sense $|c_i|\le\Delta c_i$. 

\begin{figure}[ht]
  \includegraphics[width=\hsize]{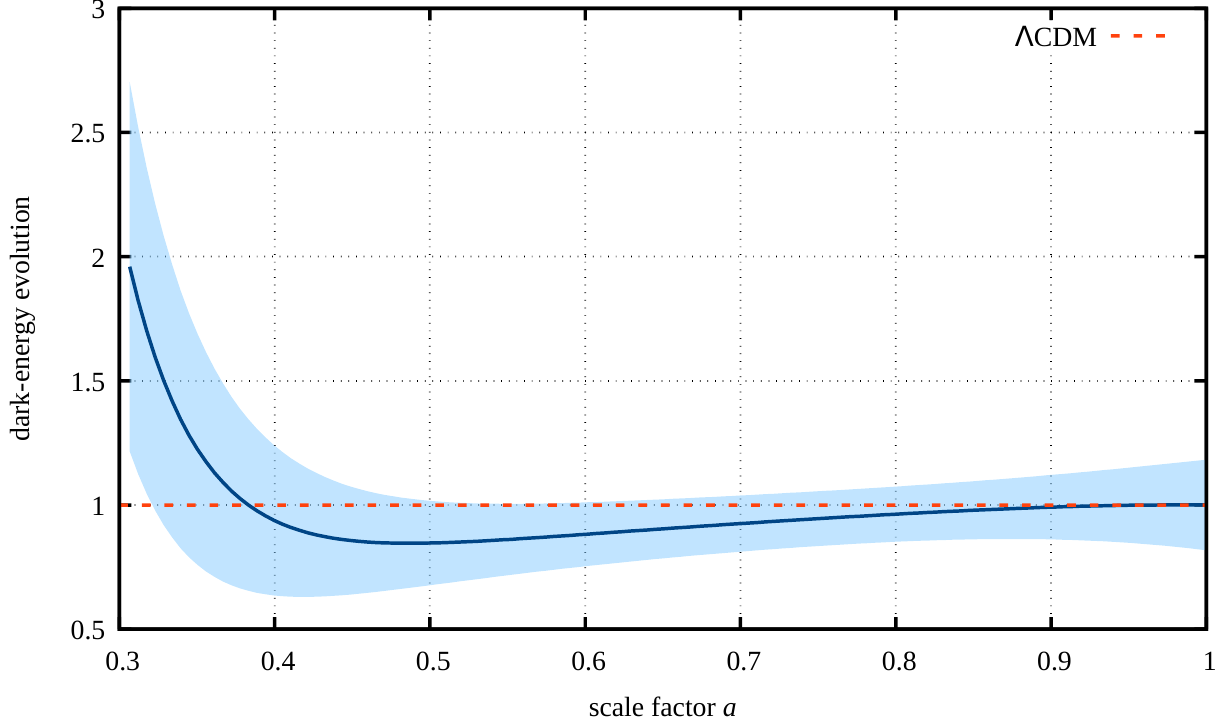}
\caption{Empirical constraint on the time evolution of the Dark Energy via the function $q(a) = [E^2(a)-\Omega_\mathrm{m0}a^{-3}]/(1-\Omega_\mathrm{m0})$ , obtained from the cosmic expansion function $E(a)$ reconstructed from the Pantheon sample of type-Ia supernovae. If Dark Energy is a cosmological constant, $q(a) = 1$. The light-blue area illustrates the 1-$\sigma$ uncertainty of $q(a)$ due to the uncertainty of the coefficients listed in Tab.\ \ref{tab:1}.}
\label{fig:1}
\end{figure}

The reconstructed expansion function $E(a)$ allows us to set limits on a possible time dependence of the Dark Energy. To do so, we introduce a function $q(a)$ by 
\begin{equation}
  E^2(a) = \Omega_\mathrm{m0}a^{-3}+(1-\Omega_\mathrm{m0})q(a)
\label{eq:q_a}
\end{equation}
such that $q(a) = 1$ for a cosmological constant. This function $q(a)$ as reconstructed from the Pantheon sample is shown in Fig.~\ref{fig:1}.

\begin{table}
  \caption{Significant expansion coefficients of the function $e(x)$ in terms of shifted Chebyshev polynomials together with their uncertainties.}
  \label{tab:1}
  \begin{tabular}{c|c|c}
    $n$ & $c_n$ & $\Delta c_n$ \\
    \hline
    $0$ &  $0.988$ & $0.033$ \\
    $1$ & $-0.372$ & $0.035$ \\
    $2$ &  $0.045$ & $0.018$ \\
  \end{tabular}
\end{table}

\subsection{Growth function}

Having reconstructed the expansion function without reference to a specific cosmological model, we can also reconstruct the growth function $D_+(a)$ of linear density perturbations in a model-independent way, again following \cite{BM2019}. The gravitational field of density fluctuations with density contrast $\delta_m = \rho/\bar\rho-1$ causes cosmic structures to grow. For small density contrast $\delta_m<1$, we can study structure growth in the linear approximation. In the standard $\Lambda$CDM model, the growth of the linear density contrast is given by
\begin{equation}
  \ddot{\delta}_m+2H\dot{\delta}_m=4\pi G \rho_m \delta_m\;.
\end{equation} 
Separating the dependences of the density contrast on space and time as $\delta_m(t,x) = D(t)f(x)$, the standard equation for the function $D(a)$ becomes
\begin{equation}
  D''+\left(\frac{3}{a}+\frac{E'(a)}{E(a)}\right)D' =
  \frac{3}{2}\frac{\Omega_\mathrm{m0}}{a^2}D \,.
\end{equation}
The linear growth factor $D_+(a)$ is the growing solution of this equation. Since we already reconstructed the cosmic expansion function $E(a)$ empirically from the distance measurements in a model independent way, we can simply determine the linear growth from it once we set the matter-density parameter at the present time, $\Omega_\mathrm{m0}$, and the initial conditions at the smallest value of the scale factor $a_\mathrm{min}$ being considered. We can choose $D_+(a_\mathrm{min}) = 1$ and $D_+'(a_\mathrm{min}) = n/a_\mathrm{min}$, with
\begin{equation}
  n = \frac{1}{4}\left[
    -1-\varepsilon+\sqrt{(1+\varepsilon)^2+24(1-\omega)}
  \right]\;.
\end{equation}
The parameters $\varepsilon$ and $\omega$ appearing here are defined by
\begin{equation}
  \varepsilon := 3+2\frac{\D\ln E}{\D\ln a}\;,\quad
  \omega := 1-\Omega_\mathrm{m}(a) = 1-\frac{\Omega_\mathrm{m0}}{E^2a^3}\;.
\end{equation} 
They are much smaller than unity during the matter-dominated epoch and can both be determined from the reconstructed expansion function once $\Omega_\mathrm{m0}$ has been set.

\begin{figure}[ht]
  \includegraphics[width=\hsize]{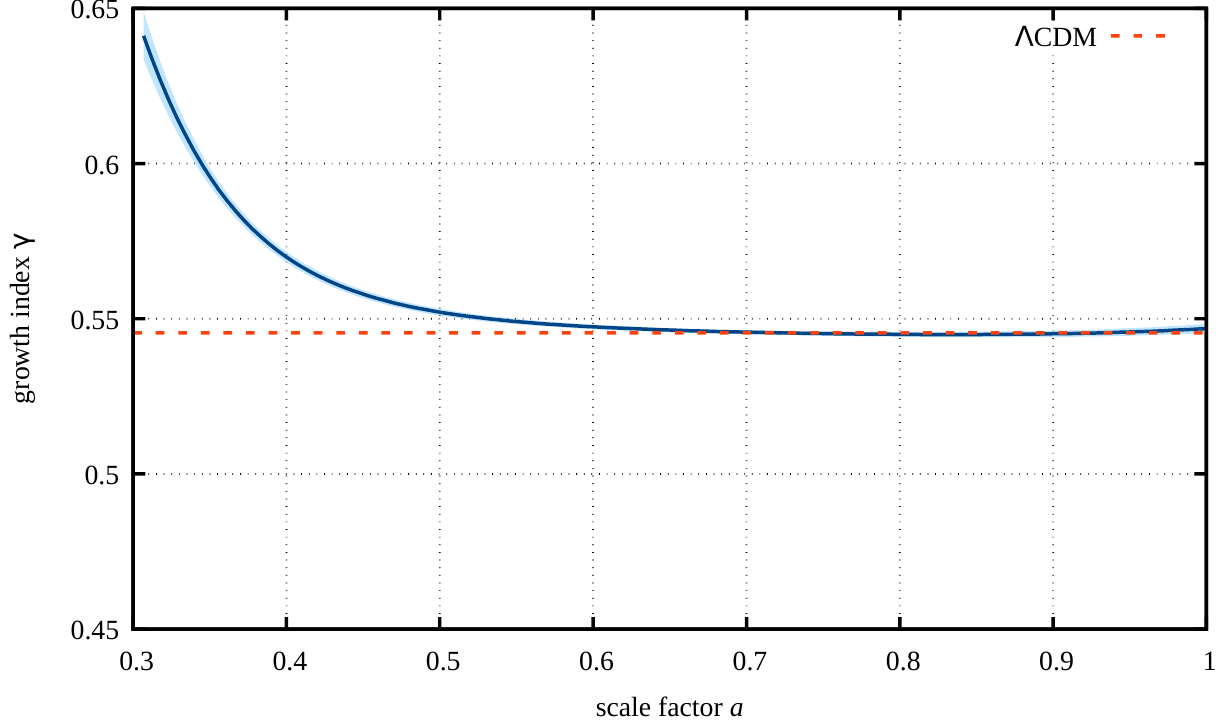}
\caption{Growth index $\gamma(a)$ for linear cosmic structures, reconstructed from the empirically determined cosmic expansion function $E(a)$. The light blue area illustrates the results for $\Omega_\mathrm{m0} = 0.3\pm0.02$. For $\Lambda$CDM, $\gamma = 6/11$ as indicated by the red-dashed line.}
\label{fig:2}
\end{figure}

We characterize the growth function $D_+(a)$ by the growth index $\gamma$ defined by
\begin{equation}
  \frac{\D\ln D_+}{\D\ln a} = \Omega_\mathrm{m}^\gamma(a)\;.
\end{equation}
As shown in \cite{BM2019}, the growth index can be written as
\begin{equation}
  \gamma = \frac{\varepsilon+3\omega}{2\varepsilon+5\omega}
\end{equation}
with
\begin{equation}
  \varepsilon := 3+2\frac{\D\ln E}{\D\ln a} \quad\mbox{and}\quad
  \omega := 1-\Omega_\mathrm{m}(a)\;.
\end{equation}
The result, determined from the empirically reconstructed expansion function $E(a)$, is shown in Fig.\ \ref{fig:2}.

As we see in Figs.\ \ref{fig:1} and \ref{fig:2}, current observations of the Pantheon supernova constrain the deviations from the standard $\Lambda$CDM model, which would mean a de Sitter attractor for the future evolution. On the other hand, both the Swampland and the TCC conjectures forbid the evolution towards a de Sitter attractor in the future. Note that the data for the growth index shows a significant deviation compared to the predictions of the vanilla $\Lambda$CDM model.

\section{Specific Dark Energy models}

\subsection{Quintessence Scalar Field}

In this section we would like to apply our model-independent analysis of the expansion and growth functions to specific Dark Energy models. One widely studied case is quintessence with an exponential potential. This case is particularly relevant for the discussion of the connection with the Swampland and TCC constraints. We begin with the action
\begin{equation}
\mathcal{S}=\int d^4x \sqrt{-g}\left\{ \frac{\mpl^2}{2}R-\frac12\partial_\mu\pi\partial^\mu\pi-V(\pi) \right\}+\mathcal{S}_{\rm matter}\,,
\end{equation}
with a quintessence field $\pi$ that admits the background field configuration $\pi=\pi(t)$. The scalar field adds the pressure $P_\pi=\frac12\dot{\pi}^2-V$ and the energy density $\rho_\pi=\frac12\dot{\pi}^2+V$. The equation of motion of the scalar field $\ddot\pi+3H\dot\pi+V'=0$ can be rewritten as a continuity equation $\dot\rho_\pi+3H(\rho_\pi+P_\pi)=0$. Einstein's field equations become
\begin{eqnarray}
3\mpl^2H^2&=&\frac{\dot\pi^2}{2}+V+\rho_m\,,\\
2\mpl^2\dot{H}&=&-\left( \dot\pi^2+(1+w_m)\rho_m \right)\;,
\end{eqnarray}
with the equation-of-state parameter of the standard matter fields $w_m$ and its energy density $\rho_m$. The density parameter of the scalar field $\Omega_\pi=\rho_\pi/(3\mpl^2H^2 )$ and the matter density parameter $\Omega_m$ satisfy $\Omega_\pi=1-\Omega_m$. The cosmological background dynamics can be brought into the form of autonomous equations \cite{Tsujikawa:2013fta}
\begin{align}\label{EOMquintessence_a}
\frac{dx}{dN}&=-3x+\frac{\sqrt{6}}{2}\lambda y^2+\frac32x\mathcal{F}\;,\\
\frac{dy}{dN}&=-\frac{\sqrt{6}}{2}\lambda xy+\frac32y\mathcal{F}\;,\\
\frac{d\lambda}{dN}&=-\sqrt{3(1+w)(x^2+y^2)}\left(\frac{VV''}{V'^2}-1\right)\lambda^2 \;,
\label{EOMquintess}
\end{align}
where 
\begin{eqnarray}
x \, &\equiv& \,  \dot\pi/(\sqrt{6}H\mpl) \, , \nonumber \\ 
y \, &\equiv& \, \sqrt{V}/(\sqrt{3}H\mpl) \, , \nonumber \\
\lambda \, &\equiv& -\mpl V'/V \, ,\\ 
N \, &\equiv& \, \ln a\, \,\,\, {\rm{and}} \nonumber \\ 
\mathcal{F} \, &\equiv& \, \left[(1-w_m)x^2+(1+w_m)(1-y^2)\right]\, . \nonumber
\end{eqnarray}
Note that $\lambda$ is a constant if we consider an exponential potential. The equation-of-state and the density parameter of the scalar field simply become 
\be
w \, = \, (x^2-y^2)/(x^2+y^2)
\end{equation}
and 
\begin{equation}
\Omega_\pi \, = \, x^2+y^2 \, , 
\end{equation}
respectively. Deep inside the Hubble radius in the quasi-static approximation the matter perturbations follow the following approximate evolution equation in the presence of the quintessence field
\begin{equation}\label{growthEqQuint}
\delta_m''+\frac12(1-2w\Omega_\pi)\delta_m'-\frac32\Omega_m \delta_m\approx 0\,.
\end{equation}

The trans-Planckian censorship conjecture in the presence of the scalar field reads
\begin{equation}\label{TCCscalar}
\int_{\pi_i}^{\pi_f}\frac{H}{\dot\pi}d\pi=
\int_{t_i}^{t_f}H dt<\ln \left(\frac{\mpl}{H_f}\right)\,.
\end{equation}
For a positive potential, this bound implies \cite{BeVa}
\begin{equation}
\frac{|\pi_f-\pi_i|}{\sqrt{(d-1)(d-2)}} < -\ln \left(\frac{H_f}{M_{pl}} \right) \,.
\end{equation} 
Since in the Friedmann equation the kinetic term of the scalar field is positive as well we can translate this upper bound into a bound for the potential
\begin{equation}\label{TCCscalarPot}
V<\frac{(d-1)(d-2)}{2}e^{-\frac{2}{\sqrt{(d-1)(d-2)}}|\pi_f-\pi_i|}.
\end{equation}
The condition \eqref{TCCscalar} or similarly its form \eqref{TCCscalarPot} is not tightly constraining for any Dark Energy model for a short evolution time due to large difference between $\mpl$ and $H_f$. However, if one considers the integral in \eqref{TCCscalar} when applied to an infinite evolution time or equivalently over an infinitely large field values, i.e. in the limit $\pi_f\to\infty$ and $\pi_i\to\infty$, this bound on the potential takes the form of the de Sitter conjecture $\frac{|V'|}{V}\ge c\sim \mathcal{O}(1)$, where $c=\lambda$ takes the specific value \cite{BeVa}
\begin{equation}\label{TCCscalarLargePi}
\frac{|V'|}{V}\ge \frac{2}{\sqrt{(d-1)(d-2)}}\,.
\end{equation}
Thus, in four dimensions this condition becomes $\frac{|V'|}{V}\ge\sqrt{2/3}\sim0.82$.

\begin{figure}[h!]
  \includegraphics[width=\hsize]{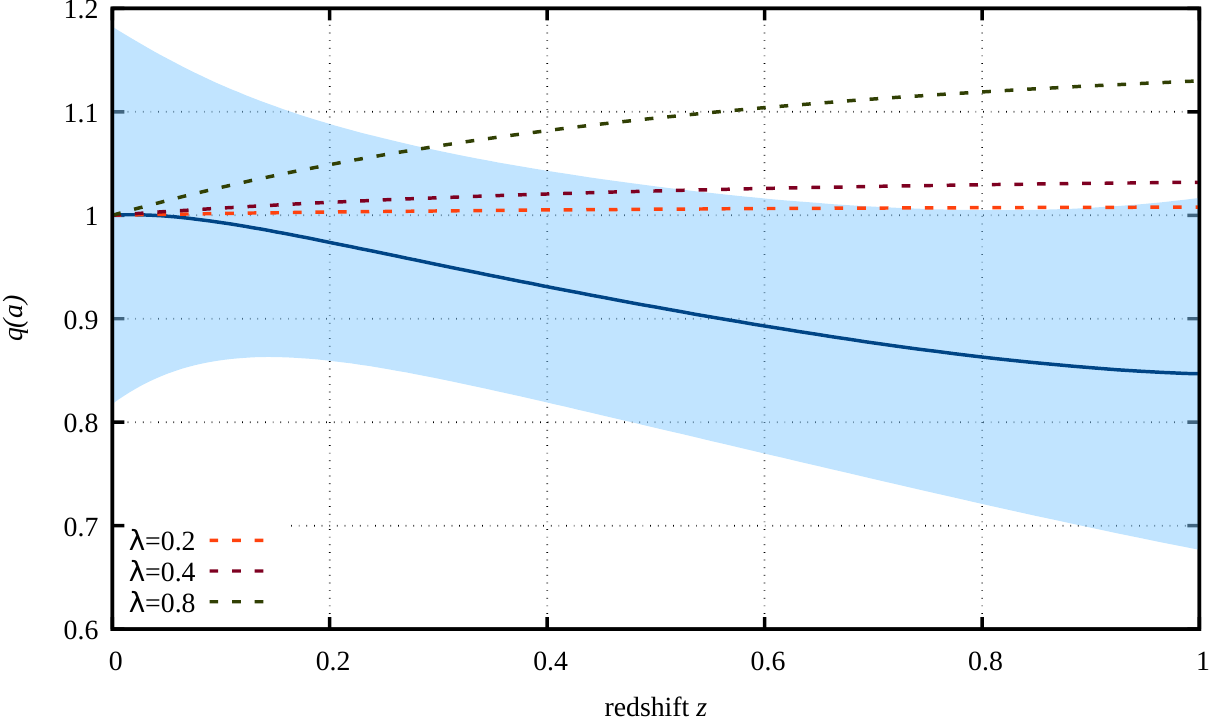}
  \includegraphics[width=\hsize]{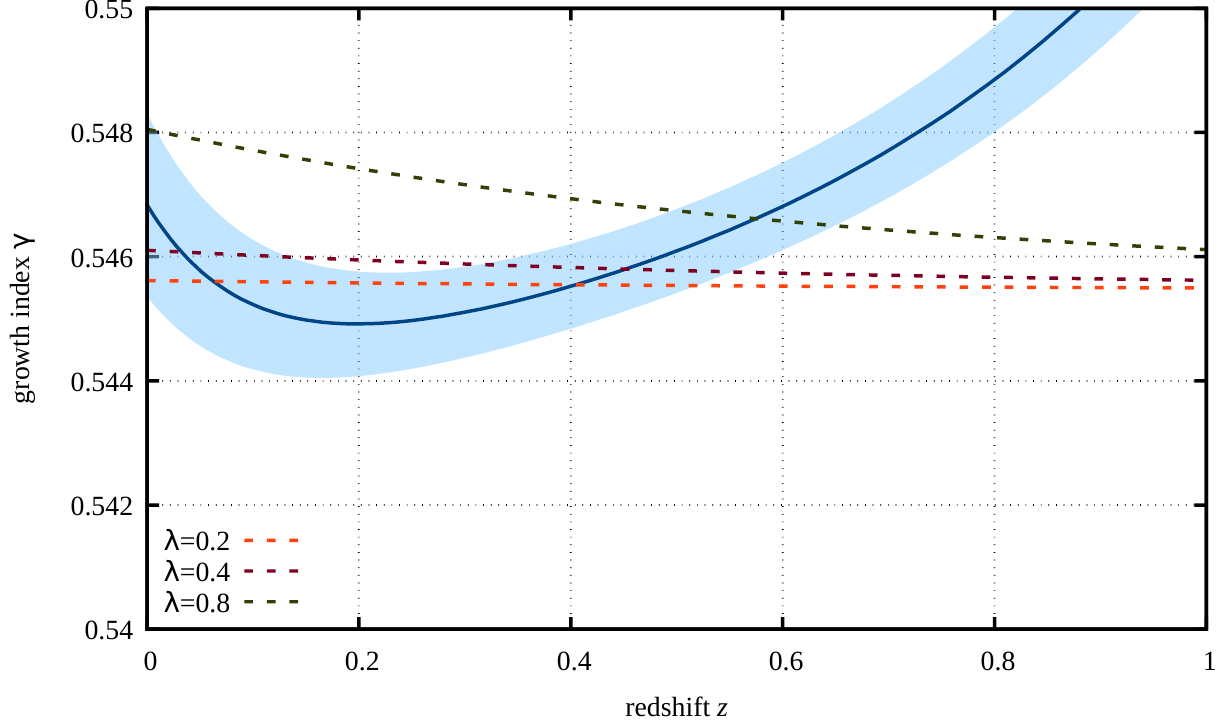}
\caption{Top: Comparison of the expected evolution of the dark-energy density with redshift for different values of $\lambda$ with the empirical constraints from the Pantheon sample. Bottom: The same for the growth index $\gamma$. The error bands show the 1-$\sigma$ uncertainty. The bottom panel shows that even small values of $\lambda$ conflict with current observational constraints on the growth function at early times.}
\label{fig_euclid}
\end{figure}

Thus, in this quintessence Dark Energy model, one has the following situation:
\begin{itemize}
\item If one focuses on the time (and hence field) interval where Dark Energy is observed to be present, then, due to the large hierarchy between the scales $\mpl$ and $H_f$ the TCC constraint \eqref{TCCscalar} is very easily satisfied and hence the TCC constraint does not have any constraining power on present observables.
\item However, if one assumes that the evolution of the quintessence field is decribed to the infinite future with the same potential, then the TCC constraint becomes a very strong constraint on the value of $\lambda$. However, in this case one would be extending the range of field values beyond what is admissible according to the distance conjecture.
\end{itemize}

\begin{figure}[h!]
  \includegraphics[width=0.8\hsize]{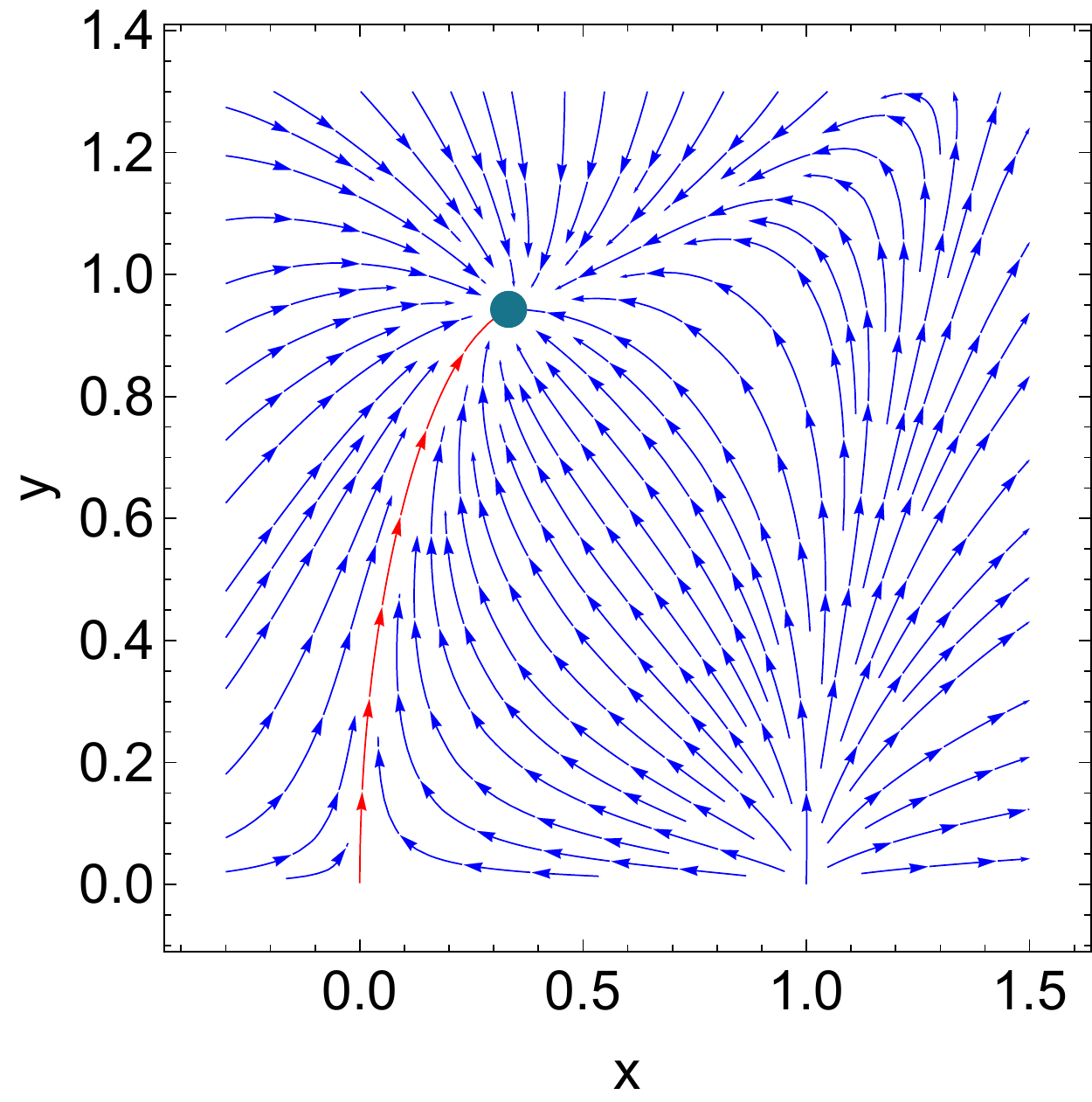}
\caption{The phase map illustrates the possible trajectories for $\lambda=\sqrt{2/3}$ of the quintessence model. There is a global attractor point $[x=\lambda/\sqrt{6},y=\sqrt{1-\lambda^2/6}]$ with $w > -1$. The red line shows one example trajectory where the universe undergoes a radiation and matter domination phase.}
\label{fig_phaseQuint}
\end{figure}

As the trajectories in \ref{fig_phaseQuint} show, the quintessence exponential model has an attractive fixed point with $w > -1$ (the green dot in figure  \ref{fig_phaseQuint}). In the limit $\lambda \rightarrow 0$ the fixed point approaches de Sitter. This fixed point behaviour has been studied comprehensively in \cite{Copeland}. At the fixed point, to leading order in $\lambda$
\be
w \, = \, - \frac{1 - \lambda^2/6}{1 + \lambda^2/6} \, .
\ee
From the point of view of the TCC constraint, the model would be safe
in the future evolution for $\lambda \ge 2 / \sqrt{(d - 1)(d - 2)}$, since the attractive fixed point is not a de Sitter critical point. At the attractor point one has $\Omega_\phi=1$ with $[x=\lambda/\sqrt{6},y=\sqrt{1-\lambda^2/6}]$ and this is shifted from the would-be a de Sitter point at $[x=0, y=1]$. However, as we have seen in figure \ref{fig_euclid}, even small values of $\lambda$ are in conflict with current observational constraints.

\subsection{Generalized Proca Vector Field}

Apart from scalar fields, vector fields can also play an important role in the Dark Energy phenomenology. Specially, in the context of the $H_0$ tension, vector fields can naturally connect early evolution of the Hubble function with the late-time evolution and hence reduce the $H_0$ tension \cite{DeFelice:2020sdq}.
Therefore, in this subsection we will concentrate on Dark Energy models based on generalized Proca theories. 
We will consider the most general vector-tensor theories with second order equations of motion with three propagating degrees of freedom, given by 
\begin{equation}
  S = \int\D^4x \sqrt{-g} \left( {\cal L}
  +{\cal L}_M \right)\;,\quad
  {\cal L}=\sum_{i=2}^{6} {\cal L}_i\;,
\label{LagProca}
\end{equation}
with matter Lagrangian ${\cal L}_M$ and the Lagrangian densities of the generalized Proca action \cite{Heisenberg:2014rta, Jimenez:2016isa}
\begin{align}
\label{gen_Proca}
  {\cal L}_2 &= G_2(X,F,Y)\;, \\
  {\cal L}_3 &= G_3(X) [K]\;, \\
  {\cal L}_4 &= G_4(X)R+G_{4,X} \left[ [K]^2 -[K^2] \right]\;, \\
  {\cal L}_5 &= G_{5}(X) G_{\mu \nu}K^{\mu\nu}-
    \frac{G_{5,X}}{6}\left[[K]^2-3[K][K^2]+2[K^3]\right] \nonumber \\
    &-g_5(X) \tilde{F}^{\alpha\mu}\tilde{F}^\beta_{\phantom{\beta}\mu}
    K_{\alpha\beta}\;, \\
  {\cal L}_6 &= G_6(X) L^{\mu \nu \alpha \beta}
    K_{\mu\nu} K_{\alpha\beta}
    +\frac{G_{6,X}}{2}\tilde{F}^{\alpha \beta} \tilde{F}^{\mu \nu}
    K_{\alpha\mu} K_{\beta\nu}\;.
\end{align}
The field strength and its dual are denoted by $F_{\mu \nu}=\nabla_{\mu}A_{\nu}-\nabla_{\nu}A_{\mu}$ and $\tilde{F}^{\mu \nu}=\epsilon^{\mu \nu \alpha \beta}F_{\alpha \beta}/2$. The operator $K$ stands for $K_{\mu\nu}=\nabla_{\mu}A_{\nu}$ and $G_{i,X} \equiv \partial G_{i}/\partial X$. While the quadratic Lagrangian ${\cal L}_2$ depends on  
$X =-\frac12 A_{\mu} A^{\mu}$, 
$F = -\frac14 F_{\mu \nu} F^{\mu \nu}$ and 
$Y = A^{\mu}A^{\nu} {F_{\mu}}^{\alpha}F_{\nu \alpha}$,
the functions $G_{3,4,5,6}$ and $g_5$ can only depend on $X$. The double dual Riemann tensor follows the notation
$L^{\mu \nu \alpha \beta}=\frac14 \epsilon^{\mu \nu \rho \sigma} \epsilon^{\alpha \beta \gamma \delta} R_{\rho \sigma \gamma \delta}$. The compatibility with homogeneity and isotropy allows the background field configuration of the vector field $A$ to be $A^\mu=(A^0(t),0,0,0)$ with $A^0 = \phi(t)/N(t)$. The background equations are given by \eqref{background_eom} where the effective energy density and the pressure take the specific form \cite{DeFelice:2016yws}
\begin{align}
  \rho_{\rm DE} &=-G_2+G_{2,X}\phi^2+3G_{3,X}\phi^3+
  12G_{4,X}H^2\phi^2-6g_4H^2
  \nonumber\\ &+6G_{4,XX}\phi^2H^2\phi^2-
  G_{5,XX}H^3\phi^5-5G_{5,X}H^3\phi^3\;,\nonumber\\
  P_{\rm DE} &= G_2-G_{3,X}\phi^2\dot\phi-
  2G_{4,X}\phi\left(3H^2\phi+2H\dot\phi+2\dot H\phi\right)
  \nonumber\\ &-
  4G_{4,XX}H\phi^3\dot\phi+G_{5,XX}H^2\phi^4\dot\phi+2g_4(3H^2+\dot{H})\nonumber\\
  &+G_{5,X}H\phi^2\left(2\dot H\phi+2H^2\phi+3H\dot\phi\right)\;.
\label{eq:9}
\end{align}
Similarly, the vector field equation gives rise to the following algebraic equation for $\phi$
\begin{align}
&\phi\,\Bigl\{
    G_{2,X}+3HG_{3,X}\phi+6H^2\left(G_{4,X}+G_{4,XX}\phi^2\right)
    \nonumber\\ &-
    H^3\left(3G_{5,X}+G_{5,XX}\phi^2\right)\phi
  \Bigr\} = 0\;.
\label{eq:7}
\end{align}
For a specific Dark Energy model where the general functions follow the ansatz 
$  G_2 = b_2X^{p_2}+F$, 
  $G_3 = b_3X^{p_3}$,
  $G_4 = \frac{M_\mathrm{Pl}^2}{2}+b_4X^{p_4}$ and
$  G_5 = b_5X^{p_5}$,
the background equations simply become in this case
\begin{eqnarray}\label{eqsVT_DE}
\Omega'_\mathrm{DE}&=&\frac{(1+s)\Omega_\mathrm{DE}(3+\Omega_\mathrm{r}-3\Omega_\mathrm{DE})}{1+s\Omega_\mathrm{DE}}\nonumber\\
\Omega'_\mathrm{r}&=&-\frac{\Omega_\mathrm{r}(1-\Omega_\mathrm{r}+(3+4s)\Omega_\mathrm{DE})}{1+s\Omega_\mathrm{DE}}\,.
\end{eqnarray}
Integration of these equations yields
 \begin{equation}
  \frac{\Omega_\mathrm{DE}}{\Omega_\mathrm{DE,0}} =
  \frac{\Omega_\mathrm{r}}{\Omega_\mathrm{r,0}}\,a^{4(1+s)}\;,
\label{eq:22}
\end{equation}
where $s := p_2/p$ represents the associated Dark Energy parameter. In order to guarantee the propagation speed of gravitational waves to be luminal, we will impose $b_5=0$ and $b_4=0$. Note that for positive values of $s$, the scaling of Dark Energy is phantom-like, i.e. the total Dark Energy increases with time. This is opposite to that what happens in quintessence models where the total Dark Energy decreases in time.

The growth function on the other hand turns into
\begin{equation}\label{eomdelta}
  \delta''+\frac{1+(3+4s)\Omega_\mathrm{DE}}{2(1+s)\Omega_\mathrm{DE}}\delta'-
  \frac{3}{2}\frac{G_\mathrm{eff}}{G}(1-\Omega_\mathrm{DE})\delta=0\;,
\end{equation}
where the effective gravitational constant modifies into
\begin{equation}
  G_\mathrm{eff} = \frac{H}{4\pi\phi}\frac{\mu_2\mu_3-\mu_1\mu_4}{\mu_5}
\label{eq:13}
\end{equation}
with the exact expressions for the functions $\mu_i$ to be found in Eqs.\ (5.10)-(5.14) and (5.18) in \cite{DeFelice:2016uil}.

\begin{figure}[h!]
  \includegraphics[width=\hsize]{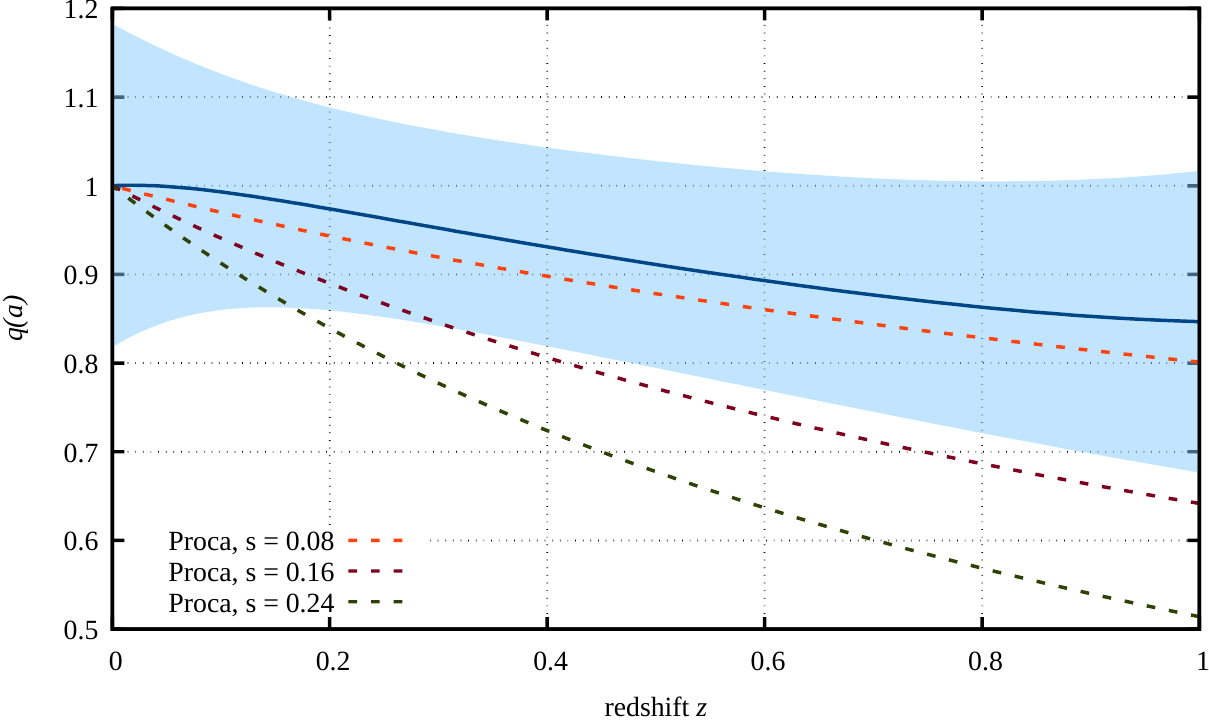}
  \includegraphics[width=\hsize]{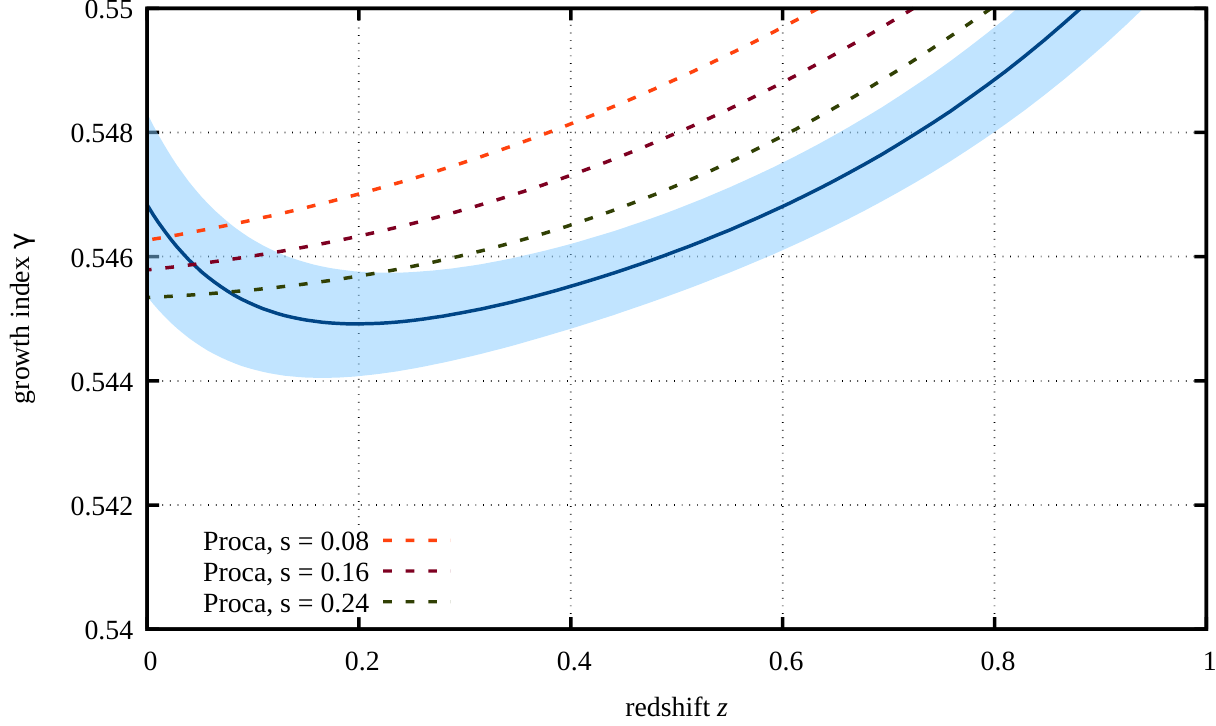}
\caption{Top: the expected evolution of the dark-energy density with redshift for different values of $s$ in the cubic Generalized Proca theory is compared with the empirical constraints from the Pantheon sample. Bottom: The same for the growth index $\gamma$. The error bands show the 1-$\sigma$ uncertainty. Small positive values of $s$ are in good agreement with current observational constraints.}
\label{fig_cubicGenProca}
\end{figure}

We apply the model-independent empirical constraints that we obtained from the Pantheon sample in terms of the shifted Chebyshev polynomials to the cubic Generalized Proca theory, which is shown in \ref{fig_cubicGenProca}. As one can see, for the appropriate values of the Dark Energy parameter $s$ we are in good agreement with current observations. It is quite intriguing that the growth index of the cubic Proca has the same trend as a function of redshift as the model-independent reconstructed growth index from the shifted Chebyshev polynomials. This is a consequence of the phantom-like scaling of Dark Energy. Previous studies have shown that phantom Dark Energy is marginally preferred by the data \cite{phantom}.

The dynamical system of the background equations of motion has three critical points $(\Omega_\mathrm{DE},\Omega_\mathrm{r})=(0,1)$, $(\Omega_\mathrm{DE},\Omega_\mathrm{r})=(0,0)$ and $(\Omega_\mathrm{DE},\Omega_\mathrm{r})=(1,0)$, where the first represents a radiation point, the second a matter and the third is the phantom dark energy fixed point with an equation of state $w < -1$, tending to $w = -1$ as $s$ decreases to zero. This fixed point is not consistent with the TCC. 

\begin{figure}[h!]
  \includegraphics[width=0.8\hsize]{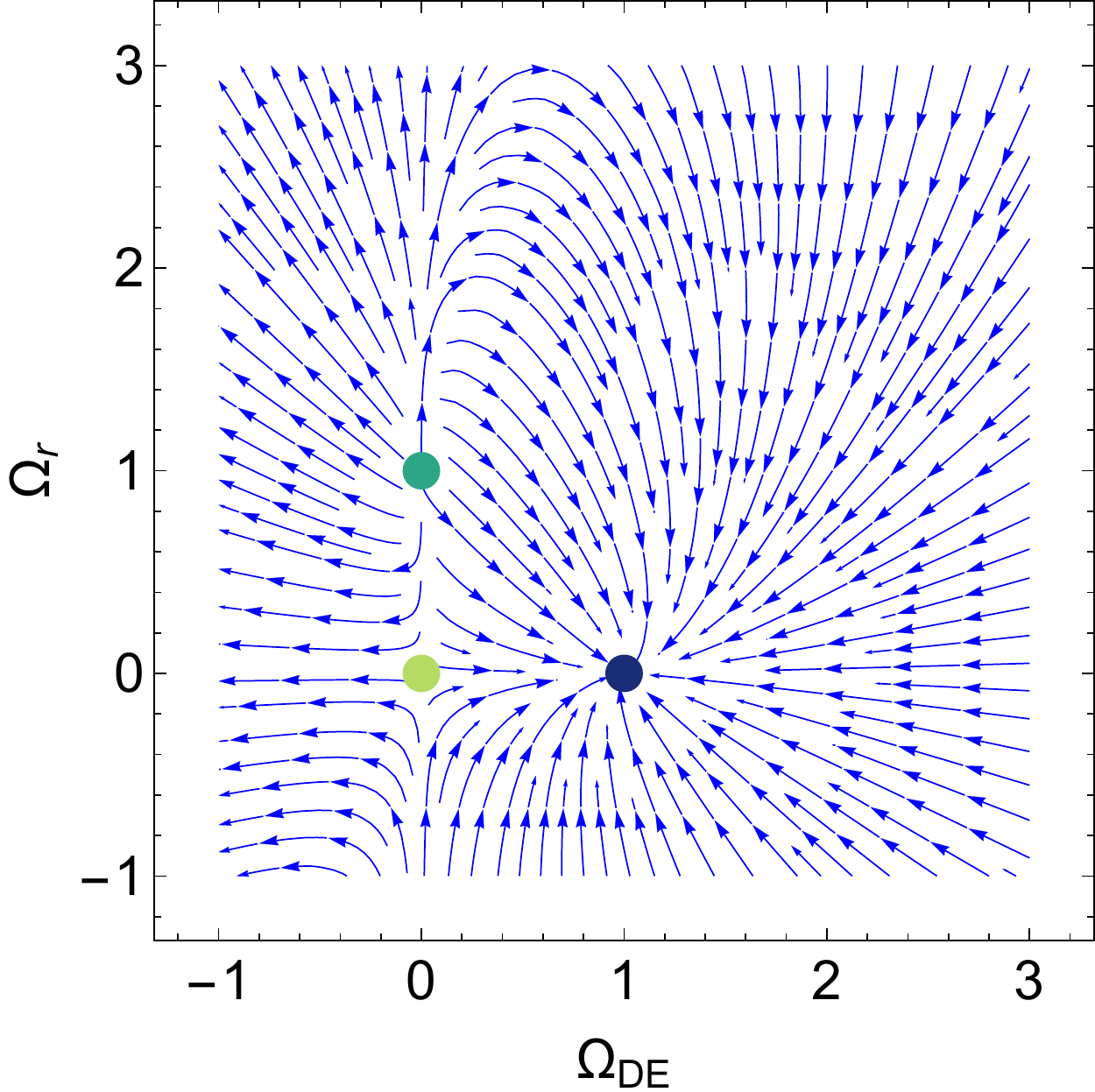}
\caption{A phase-map of the autonomous system of the Dark Energy models based on generalized Proca with the general functions given by power laws. The three critical points $(\Omega_\mathrm{DE},\Omega_\mathrm{r})=(0,1)$, $(\Omega_\mathrm{DE},\Omega_\mathrm{r})=(0,0)$ and $(\Omega_\mathrm{DE},\Omega_\mathrm{r})=(1,0)$ are independent of $s$, where for the plotted specific phase-map we have assumed $s=0.1$. The critical point $(\Omega_\mathrm{DE},\Omega_\mathrm{r})=(1,0)$ corresponds to a phantom-like Dark Energy fixed point with an equation of state $w < -1$.}
\label{fig_VTmap}
\end{figure}

In Dark Energy models given by \eqref{eqsVT_DE}, as can be seen in figure \ref{fig_VTmap}, all trajectories starting from a radiation or matter dominated epoch will end in an attractive critical point  (the blue dot in figure \ref{fig_VTmap}) which has $w < -1$ and hence is in even stronger conflict with the TCC than the vanilla $\Lambda$CDM model. If these models discussed here indeed describe the late time cosmology (to arbitrarily late times), then the TCC would be false. However, this would imply that quantum fluctuations smaller than the Planck scale will at some point in time exit the Hubble radius, leading to a severe unitarity problem in the framework of effective field theory. It is interesting that these models appear to be in better agreement with current supernovael observations (see figure \ref{fig_cubicGenProca} and \cite{deFelice:2017paw}) than standard quintessence models.  

\section{Conclusion}

We have compared the predictions of two classes of Dark Energy models with a model-independent reconstruction of the expansion and growth functions based on the Pantheon supernova data set. The vanilla $\Lambda$CDM model is in good agreement with the data, except for a disagreement concerning the growth function at early times. Quintessence models generally make the disagreement worse, whereas the generalized Proca vector field models with $s > 0$ are in better agreement with the time evolution of the growth function. This is due to the fact that these Proca vector field models have a phantom-like late time fixed point with $w < -1$, whereas quintessence models have a late time fixed point with $w > -1$. 

If the fixed points in our Dark Energy models can be trusted to arbitrarily late times, then the models which are in agreement with the data we have analyzed would be in conflict with the Trans-Planckian censorship conjecture. However, from the point of view of string theory one should not trust the extrapolation of a quintessence model to arbitrarily large field values, since doing so would lead to a conflict with one of the other Swampland criteria, the distance conjecture. Based on the considerations which lead to the distance conjecture, one would expect the scalar field model to exit the range of field values for which it is valid after having moved a field distance of the order one in Planck units. After that point, the description of the dynamics in terms of the original quintessence field would break down.

The phantom behaviour of the effective field theory which the data seems to mildly prefer may arise naturally from considerations of the Swampland, as recently discussed in \cite{Vafa3}. The breakdown of the effective field theory of dark energy is tied to the fact that masses of other particles depend on the value of the Dark Energy field. If these other particles include the dark matter particles, then one would expect a drift of energy from the dark matter sector to the dark energy sector as the dark matter particles become lighter. This would result in an effective phantom equation of state for the dark energy field.

In this work we have represented Dark Energy models with an effective energy density and pressure. Whatever is the source for this Dark Energy component, we kept the specific form of the expansion rate arbitrary without defining a cosmological model a priori. Instead we reconstructed the expansion function directly from the Pantheon supernova sample in a model-independent way. For this purpose we expanded the Hubble function and the luminosity distance in terms of shifted Chebyshev polynomials and determined the coefficients of the polynomials by the maximum-likelihood fit from data. In a similar way, from the cosmic expansion rate we also reconstruct the growth function of linear density perturbations in a model-independent way.

To end with a more philosophical comment: Even though building sufficiently powerful particle accelerators to test quantum gravity will be very difficult, if not impossible, cosmology comes to help. Due to observational high precision, which will improve with future surveys, cosmology is a suitable arena to test fundamental physics of gravity. Specially, the new Swampland framework brings cosmology and quantum gravity closer than it has ever been. The Swampland criteria have very important consequences, not only for construction of consistent effective field theories but also for cosmology. The candidates for fundamental theory of quantum gravity needs to be reconcilable with the observed Dark Energy in the universe.


\section*{Acknowledgements}
LH is supported by funding from the European Research Council (ERC) under the European Unions Horizon 2020 research and innovation programme grant agreement No 801781 and by the Swiss National Science Foundation grant 179740. RB is supported in part by the Canadian NSERC and by the Canada Research Chairs program. He is grateful to the Institutes for Particle Physics and Astrophysics and Theoretical Physics for hospitality at the ETH.


\end{document}